\documentclass[aps,prl,preprint,groupedaddress,showpacs]{revtex4-1}
 \usepackage[utf8]{inputenc}
 \usepackage[T1]{fontenc}
 \usepackage{amsmath}
  \usepackage{amsfonts}
 \usepackage{amssymb}
 \usepackage{amsthm}
 \usepackage{array}
 \usepackage{booktabs}
 \usepackage[pdftex]{hyperref}
 \usepackage{cancel}
 \usepackage{graphicx}
 \usepackage{float}
 \usepackage{subcaption}

 \newcommand{\bra}[1]{\langle #1|}
\newcommand{\ket}[1]{|#1\rangle}

\DeclareMathOperator{\affone}{\lower-0.5ex\hbox{\normalsize$^1$}}
\DeclareMathOperator{\afftwo}{\lower-0.5ex\hbox{\normalsize$^2$}}
\DeclareMathOperator{\affthree}{\lower-0.5ex\hbox{\normalsize$^3$}}

\DeclareMathOperator{\affpoint}{\lower-0.5ex\hbox{\normalsize$^,$}}

\begin{document}
\raggedbottom
\preprint{HIP-2016-03/TH} 

\title{Constraining the non-standard interaction parameters in long baseline neutrino experiments}

\author{K. Huitu\footnote{e-mail: katri.huitu@helsinki.fi}}
\affiliation{Department of Physics, and Helsinki Institute of Physics, P. O. Box 64, FIN-00014 University of Helsinki, Finland}
\author{T. J. Kärkkäinen\footnote{e-mail: timo.j.karkkainen@helsinki.fi}}
\affiliation{Department of Physics, University of Helsinki, P.O.Box 64, FI-00014 University of Helsinki, Finland}
\author{J. Maalampi\footnote{e-mail: jukka.maalampi@jyu.fi} and S. Vihonen\footnote{e-mail: sampsa.p.vihonen@student.jyu.fi}}
\affiliation{University of Jyvaskyla, Department of Physics, P.O. Box 35, FI-40014 University of Jyvaskyla, Finland}

\date{\today}

\begin{abstract}
In this article we investigate the prospects for probing the strength of the possible non-standard neutrino interactions (NSI) in long baseline neutrino oscillation experiments. We find that these experiments are sensitive to NSI couplings down to the level of 0.01-0.1 depending on the oscillation channel and the baseline length, as well as on the detector's fiducial mass. We also investigate the interference of the leptonic CP angle $\delta_{CP}$ with the constraining of the NSI couplings. It is found that the interference is strong in the case of  the $\nu_{e}\leftrightarrow\nu_{\mu}$ and $\nu_{e}\leftrightarrow\nu_{\tau}$ transitions but not significant in other transitions. In our numerical analysis we apply the GLoBES software and use the LBNO setup as our benchmark.
\end{abstract}


\maketitle

\section{Introduction}

The discovery of neutrino oscillation and neutrino flavour conversion belongs to the major achievements of particle physics in past few decades. The observation of the oscillation of atmospheric neutrinos by the Super-Kamiokande experiment  \cite{Jung:2001dh, Kajita:2000mr} and the flux measurements of the solar neutrinos by the SNO experiment  \cite{Ahmad:2001an} and the earlier solar neutrino experiments \cite{Davis:1968cp}, firmly establish that neutrinos  are massive particles and lepton flavours mix.  The parameters describing neutrino masses and flavour mixing  have been extensively studied in many atmospheric, solar, accelerator and reactor neutrino experiments, resulting in high-precision constraints on their values, see e.g. Refs. \cite{Wendell:2010md,Adamson:2013whj,Abe:2014ugx,Adamson:2013ue,Abe:2013hdq}.  The present experimental values of the oscillation parameters are presented in Table \ref{para1}. There still are, however, some unknowns in the neutrino mixing scheme of the standard three neutrinos. One of them is the question of mass hierarchy, namely does there exist  two light neutrinos  and one heavier neutrino (the normal hierarchy, NH) or one light neutrino and two heavier ones (the inverted hierarchy, IH). Also, the question of possible CP violation in the lepton sector is still open, as is the octant of the leptonic mixing angle $\theta_{23}$, too. All these open questions will be adressed in future experimental studies, such as the long baseline neutrino oscillation experiments which have been under discussions recently, see e.g. \cite{Stahl:2012exa, Agostino:2014qoa, Bora:2015lyo, Ahmed:2015jtv, An:2015jdp, Das:2014fja}. 

The masses of neutrinos and neutrino flavour mixing cannot be explained in the framework of the minimal Standard Model  (SM), since in the SM neutrinos are considered to be massless particles. The observation of neutrino oscillations and flavour conversion thus indicate that there is some physics beyond the SM. Besides being responsible for neutrino masses and mixing, this new physics may manifest itself also as new interactions affecting the processes through which neutrinos are created and detected, and they can also give rise to new effects on neutrinos propagating in matter. Such new interactions are called non-standard neutrino interactions (NSI). 

The possible impacts of NSI have been widely studied, and constraints on the parameters modelling their effects at low energies have been derived from a great variety of experimental results. For recent reviews on NSI, see Refs. \cite{Miranda:2015dra, Ohlsson:2012kf}. No definite evidence of NSI has appeared but  all observations made so far can be explained in terms of the standard interactions of the known three neutrinos, some of them with the help of sterile neutrino(s). In some cases NSI can give an alternative explanation to experimental results (see e.g. \cite{Girardi:2014kca}) but so far such explanations have never been the only possibility.

In this paper we shall investigate the potential of the future long baseline neutrino oscillation experiments for detecting NSI. In these experiments,  NSI can affect both neutrino production in the source and the detection process at detectors, as well as the neutrino propagation in the Earth's crust. We will concentrate on the matter effect in this study. We will also study the interference of the CP angle $\delta_{CP}$, whose value is still unknown, with the determination of the NSI parameters from the oscillation data.

\begin{table}
\begin{center}
\begin{tabular}{|c|c|c|}\hline
\rule{0pt}{3ex}Parameter & Value $\pm$ Error \\ \hline
\rule{0pt}{3ex}$\sin^2 \theta_{12}$ & $0.304 \pm 0.013$ \\ \hline
\rule{0pt}{3ex}$\sin^2 \theta_{13}$ & $0.0218 \pm 0.001$\\ \hline
\rule{0pt}{3ex}$\sin^2 \theta_{23}$ & $0.562 \pm 0.032$\\ \hline
\rule{0pt}{3ex}$\Delta m_{21}^2$ & $(7.500 \pm 0.019)\cdot 10^{-5}$ eV$^2$\\ \hline
\rule{0pt}{5ex}$\Delta m_{31}^2$ & $
\left( \begin{array}{c}
2.457 \pm 0.045\\
-2.449 \pm 0.048
\end{array} \right)\cdot 10^{-3}$ eV$^2$ \\ \hline
\rule{0pt}{3ex}$\delta_{CP}$ & $\left\lbrace 2\pi n/25 | n =1,\cdots ,25 \right\rbrace\pm 0$ \\ \hline
\end{tabular}
\caption{\label{para1}Standard neutrino oscillation parameters. (See Refs. \cite{Bergstrom:2015rba, Gonzalez-Garcia:2014bfa}) 
For the unknown $\delta_{CP}$ we have denoted the values considered in numerical calculations.}
\end{center}
\end{table}

\section{The basic NSI formalism}

 In the low-energy regime,  NSI can be parametrized in terms of  effective charged current like (CC) and neutral current like (NC)  Lagrangians, given respectively by 
\cite{Grossman:1995wx}
\begin{equation}\label{lnsi}
\begin{split}
&\mathcal{L}^{CC}_{NSI} = -2\sqrt{2}G_F\varepsilon_{\alpha\beta}^{ff',C}(\overline{\nu}_{\alpha}\gamma^{\mu}P_L\ell_{\beta})(\overline{f}\gamma^{\mu}P_Cf'),\\
&\mathcal{L}^{NC}_{NSI} = -2\sqrt{2}G_F\varepsilon_{\alpha\beta}^{f,C}(\overline{\nu}_{\alpha}\gamma^{\mu}P_L\nu_{\beta})(\overline{f}\gamma^{\mu}P_Cf).
\end{split}
\end{equation}
Here $f$ and $f'$ label charged leptons or quarks ($\ell_i, u_i, d_i,\, i=1,2,3$) , $G_F = 1.166\cdot 10^{-5}$ GeV$^{-2}$ is the Fermi coupling constant, $\alpha,\beta$ refer to neutrino flavour ($e,\mu,\tau$), and $C=L,R$  refers to the chirality structure of the charged lepton interaction, $P_L$ and $P_R$ being the chiral projection operators. The NSI parameters $\varepsilon_{\alpha\beta}^{ff',C}$ and $\varepsilon_{\alpha\beta}^{f,C}$ are dimensionless numbers. It is assumed here that the effective non-standard interactions have V$-$A Lorentz structure, and for the charged fermions we allow both  left-handed ($P_C=P_L$) and right-handed ($P_C=P_R$) couplings.   The charged current Lagrangian $\mathcal{L}^{CC}_{NSI}$ is relevant for the NSI effects in the source and detector, since both in the creation and detection processes involve charged fermions. The neutral current Lagrangian $\mathcal{L}^{NC}_{NSI}$ in turn is relevant for the NSI matter effects.   The effective  low-energy Lagrangians (\ref{lnsi}) are assumed to follow from some unspecified  beyond-the-standard-model theory after integrating out heavy degrees of freedom.

In the presence of NSI the neutrino states produced in a source and detected at a detector are not necessarily pure flavour states but they may consist of several flavours as NSI may be flavour non-diagonal. Furthermore, these states are not necessarily the same in the source and at the detector as  the physical processes involved may be different in these two places. We express these states in the following way: \cite{Ohlsson:2012kf}
\begin{equation}
\begin{split}
&\ket{\nu^s_{\alpha}} = \ket{\nu_{\alpha}}+\varepsilon_{\alpha\beta}^s\ket{\nu_{\beta}},\\
&\bra{\nu^d_{\beta}} = \bra{\nu_{\beta}}+\varepsilon_{\alpha\beta}^d\bra{\nu_{\alpha}},
\end{split}
\end{equation}
where the superscripts $s$ and $d$ refer to the source and the detector, respectively. The matrices $\varepsilon^s$ and $\varepsilon^d$, which parameterize the effect of NSI, are in general different. Their elements are defined through the parameters $\varepsilon_{\alpha\beta}^{ff',C}$ appearing in the CC Lagrangian given in Eq.~(\ref{lnsi}), depending on the processes the neutrino production and detection are based on. 

The current experimental upper bounds for the source and detection NSI parameters $\varepsilon^{s,d}_{\alpha\beta}$ are given e.g.  in \cite{Biggio:2009nt}, and they range from 0.013 to 0.078.
In the following we will assume that $\varepsilon^{s,d}_{\alpha\beta}$ vanish, in other words, we assume that the effects of NSI on the production and detection of neutrinos is negligible and concentrate on the possible NSI effects on neutrinos  propagating in matter.

Concerning the propagation in matter, NSI could contribute to the coherent forward scattering of neutrinos in the Earth's crust. The effective Hamiltonian describing the time evolution of a neutrino state would take the form
\begin{equation}
H=\frac{1}{2E_{\nu}}\left [U{\rm diag}(m_1^2,m_2^2,m_3^2)U^{\dagger} +{\rm diag}(A,0,0) + A\varepsilon^m \right].
\end{equation}
where $E_{\nu}$ is the energy of the neutrino.
The matrix $\varepsilon^m$ parametrizes the NSI effects (the superscript $m$ stands for ”matter”), $U$ is the ordinary neutrino mixing matrix, $m_i$ are masses of the three active  neutrinos. The elements of the the matrix $\varepsilon^m$, denoted by $\varepsilon^m_{\alpha\beta}$, $\alpha,\beta={\rm e},\mu ,\tau$, are determined by the neutral current NSI parameters $\varepsilon_{\alpha\beta}^{f,C}$  through the equation \cite{Ohlsson:2012kf}
 \begin{equation}
\varepsilon^m_{\alpha\beta}=\sum_{f,C} \varepsilon_{\alpha\beta}^{f,C}\frac{N_f}{N_e}.
\end{equation}
 The effective matter potential, including both the SM and NSI matter effects, is given by the matrix
\begin{equation}
V = A\left(
\begin{array}{ccc}
1+\varepsilon_{ee}^m & \varepsilon_{e\mu}^m & \varepsilon_{e\tau}^m \\
\varepsilon_{e\mu}^{m*} & \varepsilon_{\mu\mu}^m & \varepsilon_{\mu\tau}^m \\
\varepsilon_{e\tau}^{m*} & \varepsilon_{\mu\tau}^{m*} &\varepsilon_{\tau\tau}^m
\end{array}\right).
\label{Vnsi}
\end{equation}
The effective Hamiltonian can be then written as
\begin{equation}
H = \frac{1}{2E_{\nu}}\left[U
\left(
\begin{array}{ccc}
0 & 0 & 0 \\
0 & \Delta m_{21}^2 & 0\\
0 & 0 & \Delta m_{31}^2 
\end{array}
\right) U^{\dagger}
+V\right].
\label{Hnsi}
\end{equation}
The probability of the transition $\nu_{\alpha}^s \rightarrow \nu_{\beta}^d$ is  given by
\begin{equation}
P_{\nu_{\alpha}^s \rightarrow \nu_{\beta}^d} = \left| \bra{\nu_{\beta}^d}e^{-iHL}\ket{\nu^s_{\alpha}}\right|^2,
\label{transprob}
\end{equation}
where $L$ is the baseline length.
One can extract bounds on the NSI parameters by confronting this theoretical expression with the  results of oscillation experiments. 
The current bounds on the parameters describing the matter-induced NSI  effects, as quoted in \cite{Biggio:2009nt},  are given in Table~\ref{bounds}, in the column experimental bounds. 

\begin{table}
\begin{center}
\begin{tabular}{|c|c|c|c|c|c|}\hline
\rule{0pt}{3ex}Parameter & Experimental limit & 20 kt & 50 kt & 70 kt & 150 kt\\ \hline
\rule{0pt}{3ex}$|\varepsilon^m_{ee}|$      &  < 4.2 & < 0.28 & < 0.18 & < 0.16 & <  0.084 \\ \hline
\rule{0pt}{3ex}$|\varepsilon^m_{e\mu}|$      & < 0.33 &  < 0.040 & < 0.025 & < 0.022 & < 0.018 \\ \hline
\rule{0pt}{3ex}$|\varepsilon^m_{e\tau}|$      & < 3.0 & < 0.028  & < 0.021 & < 0.019 & < 0.015 \\ \hline
\rule{0pt}{3ex}$|\varepsilon^m_{\mu\mu}|$      & < 0.068 & < 0.10 & < 0.063 & < 0.056 & <  0.040 \\ \hline
\rule{0pt}{3ex}$|\varepsilon^m_{\mu\tau}|$      & < 0.33 & < 0.013 & < 0.008 & < 0.007 & < 0.005\\ \hline
\rule{0pt}{3ex}$|\varepsilon^m_{\tau\tau}|$      & < 21 & < 0.10 & < 0.063& < 0.056 & <  0.040 \\ \hline
\end{tabular}
\end{center}
\caption{\label{bounds}Current experimental limits of the  matter NSI parameters  \cite{Biggio:2009nt}, and the expected limits from benchmark setup with different detector masses at $\delta_{CP} = \pi/2$. All limits are at 90 $\%$ confidence limit. }
\end{table}

\section{Numerical simulations}

We study numerically how the future neutrino oscillation experiments would constrain various NSI parameters. We will concentrate here on the future long baseline neutrino experiments, using the LBNO setup with a high-intensity beam, a baseline of 2300 km and 20 kt  double-phase liquid argon detector as our benchmark, see Table \ref{para2}. The analysis is done by using the GLoBES simulation software \cite{Huber:2004ka,Huber:2007ji,Kopp:2006wp,Kopp:2007ne}.

GLoBES calculates the oscillation probabilities and the corresponding neutrino rates for any given set of oscillation parameter values. The standard set of the software simulates the neutrino propagation from source to detector and computes the standard matter interactions (SI) for the distance that neutrinos travel. A software extension then allows to perform the same calculation but also includes all matter-induced NSI effects in the propagation. The software computes $\chi^2$ distributions to compare different sets of oscillation parameter values.
We determine the 90\% CL upper bounds for $\varepsilon_{\alpha\beta}^m$ by evaluating the NSI discovery potential, that is, the sensitivity to rule out SI in favor of NSI. The non-observation of NSI then allows to set new 90\% CL limits for $\varepsilon_{\alpha\beta}^m$.

\begin{table}
\begin{center}
\begin{tabular}{|c|c|}\hline 
\rule{0pt}{3ex}Runtime ($\nu + \overline{\nu}$ years) & 5+5 \\ \hline
\rule{0pt}{3ex}LAr detector mass (kt) & 20 \\ \hline
\rule{0pt}{3ex}Neutrino beam power (MW) & 0.75\\ \hline
\rule{0pt}{3ex}POT (1/year) & $1.125\cdot 10^{20}$\\ \hline
\rule{0pt}{3ex}Baseline length (km) & 2288\\ \hline
\rule{0pt}{3ex} Energy resolution function & $0.15\sqrt{E}$\\ \hline
\rule{0pt}{3ex}Energy window (GeV)& 0 --- 10\\ \hline
\rule{0pt}{3ex}Bin width (GeV) & 0.2 \\ \hline
\rule{0pt}{3ex}Bins & 50\\ \hline
\end{tabular}
\caption{\label{para2} The benchmark values of various experimental parameters used in the numerical calculations.}
\end{center}
\end{table}

The NSI discovery potential is calculated as follows. The $\chi^2$ distributions are calculated from both SI and NSI theories. The SI value $\chi_\text{SI}^2$ is computed by assuming the standard three-neutrino mixing parameters, and setting all NSI parameters in Eq.~(\ref{Vnsi}) to zero. The NSI values $\chi_\text{NSI}^2$, on the other hand, are defined by assigning one $\varepsilon_{\alpha\beta}^m$ parameter with a non-zero value. The $\Delta \chi^2$ value is then obtained from the difference between the two $\chi^2$ values:

\begin{equation}
\Delta \chi^2 = \chi_\text{SI}^2 - \chi_\text{NSI}^2.
\label{DeltaChi2}
\end{equation}

The $90\,\%$ confidence level is hence obtained at $\Delta\chi^2 = 2.71$. We take the standard oscillation parameter values from the current best-fits as determined from global analysis of experimental data \cite{Forero:2014bxa}. 
The best-fit values and their $1\,\sigma$ errors are presented in Table \ref{para1}.

For each $\delta_{CP}$ value, we have calculated the $\Delta \chi^2$ value in a baseline range 100$-$5000 km and $\log_{10} |\varepsilon_{\alpha\beta}^m|$ range from $-3.0$ to $-0.5$. In every case, a 90 $\%$ confidence level contour is found and the results merged in a contour band. The bands in $(L,\varepsilon^m_{\alpha\beta})$-plane are plotted in Figure~\ref{fig:1}.

The vertical width of the band corrensponds to the strength of correlation between $\varepsilon^m_{\alpha\beta}$ and $\delta_{CP}$. Immediately we observe that the discovery potentials reach their maximums at $\sim 2000$ km baseline. This tells us that the LBNO setup used in our numerical studies, with the baseline of 2300 km, which is close to optimal for the detection of the neutrino mass hierarchy and the leptonic CP violation, is also suitable for the NSI studies. Note that this maximum is specific for our benchmark setup: other sources will imply the maximum to be at a different baseline. However, there are common features for all setups: for example, the discovery potential is greatly reduced at shorter baselines.
Correlation between $\delta_{CP}$ and NSI parameters is notable for $|\varepsilon^m_{e\mu}|$ and $|\varepsilon^m_{e\tau}|$. 

Other parameters have only weak or nonexistent correlation, which results in the narrow bands in Figure~\ref{fig:1}.
We find that for the cases of normal and inverted mass hierarchy, there is only little difference in the confidence limits, and thus the corresponding plots for inverted hierarchy would be almost identical. 
We have taken all $\varepsilon_{\alpha\beta}$ to be real, and kept only $\delta_{CP}$ as a CP violating phase.
Letting the off-diagonal NSI parameters to be complex would widen the bands for $|\varepsilon_{e\mu}^m|$,
$|\varepsilon_{e\tau}^m|$, and $|\varepsilon_{\mu\tau}^m|$. Some effects of non-zero $\epsilon^m_{\alpha\beta}$ phases were studied in \cite{Forero:2016}.

We have given the expected model independent bounds for NSI parameters for our benchmark setup in Table \ref{bounds}. Increasing the neutrino energy for this baseline does not increase significantly the sensitivity for matter NSI, as was observed in \cite{Adhikari:2012wn}. We have studied the effect of increasing the LAr detector mass, while keeping all the other parameters of the benchmark setup as previously. In addition to 20 kt, the expected bounds have been given also for several other detector masses. 
Comparing with the present experimental bounds for the  NSI parameters \cite{Biggio:2009nt}, see Table \ref{bounds}, it is seen that with the benchmark setup, it is possible to significantly improve constraints for several NSI parameters. 

However, one should notice that for large detector masses, the limits on various NSI parameters become so stringent that the detector and source NSI parameter bounds are of the same order, and for a precise limit those should be considered as well. Thus the bounds given for larger detector masses are only indicative.

\section{Correlation between the CP angle  and NSI parameters}\label{correl}

Let us now investigate analytically the correlation between the CP angle  $\delta$ and the various NSI parameters  $\varepsilon_{\ell\ell'}^m$ in the  transition probabilities  $P(\nu_{\ell} \rightarrow \nu_{\ell'})\equiv P_{\ell\ell'}$, where  $\ell,\ell' = e,\mu,\tau$. 

The transition amplitude for $\nu_{\ell}\to\nu_{\ell'}$ is given by 
\begin{equation}
A(\nu_{\ell} \rightarrow \nu_{\ell'})\equiv A_{\ell\ell'}= \sum \limits_{j,k=1}^3\frac{L}{2E_{\nu}}U_{\ell j}H_{jk}(U^{\dagger})_{k\ell'}
\end{equation}
and the transition probability correspondingly by
\begin{equation}
P_{\ell\ell'}=\vert A_{\ell\ell'} \vert^2=\vert \sum \limits_{j,k=1}^3\frac{L}{2E_{\nu}}U_{\ell j}H_{jk}(U^{\dagger})_{k\ell'}\vert^2.
\end{equation}
We keep all the oscillation angles $\theta_{ij}$ and mass squared differences $\Delta m^2_{ij}$ fixed to their best fit values, given in Table~\ref{para1}, and treat the CP-violating phase $\delta = \delta_{CP}$ and the NSI parameters $\varepsilon^m_{\ell\ell'}$, the latter one at the time,  as the only variables.

The approximative expressions for the amplitudes $A_{\ell\ell'}$ are in general of the form
\begin{equation}\label{amp-app}
\frac{2E_{\nu}}{L}A_{\ell\ell'}=N_{\ell\ell'}+K_{\ell\ell'}e^{-i\delta}
\end{equation}
for appearance processes ($\ell\neq \ell'$) and 
\begin{equation}\label{amp-disapp}
\frac{2E_{\nu}}{L}A_{\ell\ell}=N_{\ell\ell}+K_{\ell\ell}\cos \delta 
\end{equation}
for disappearance processes. 
The expressions are accurate to first power in $s_{13}$ and $\varepsilon^m_{\ell\ell'}$.
An exception to these rules is the transition $\nu_{\mu} \rightarrow \nu_{\tau}$, for which we obtain
\begin{equation}\label{mutau}
\frac{2E_{\nu}}{L}A_{\mu\tau} = N_{\mu\tau}+K_{\mu\tau}^{(-)}e^{-i\delta}+K_{\mu\tau}^{(c)}\cos\delta.
\end{equation}
In Eqs. \eqref{amp-app},  \eqref{amp-disapp} and \eqref{mutau} the dependence on the NSI parameters and neutrino mixing angles and squared mass differences are given in terms of  the functions $N_{\ell\ell}, N_{\ell\ell'}, K_{\ell\ell}$ and $K_{\ell\ell'}$, which are defined in Table~\ref{NK}. 
In these functions, the relevant NSI parameter $\varepsilon^m_{\ell\ell}$ or $\varepsilon^m_{\ell\ell'}$, as well as $\sin \theta_{31}$,  are taken into account to the first order. Similar expressions are calculated in \cite{Kopp:2007ne}.

\begin{figure}[H]
\begin{subfigure}{0.41\textwidth}
\includegraphics[width=\linewidth]{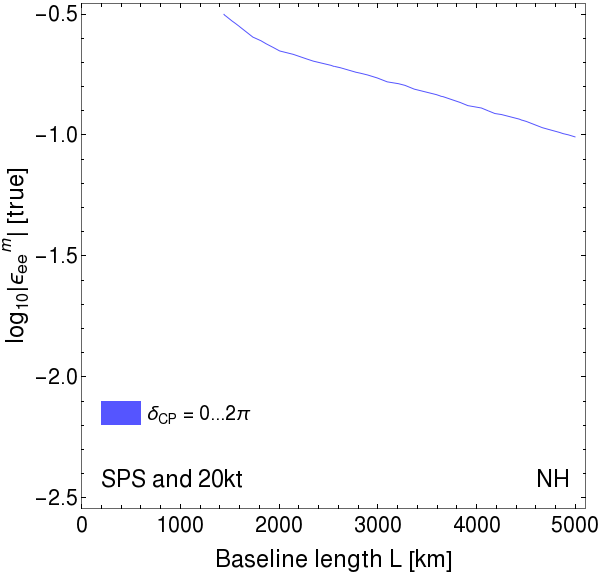}
\caption{Case $\varepsilon^m_{ee}$} \label{fig:a}
\end{subfigure}\hspace*{\fill}
\begin{subfigure}{0.41\textwidth}
\includegraphics[width=\linewidth]{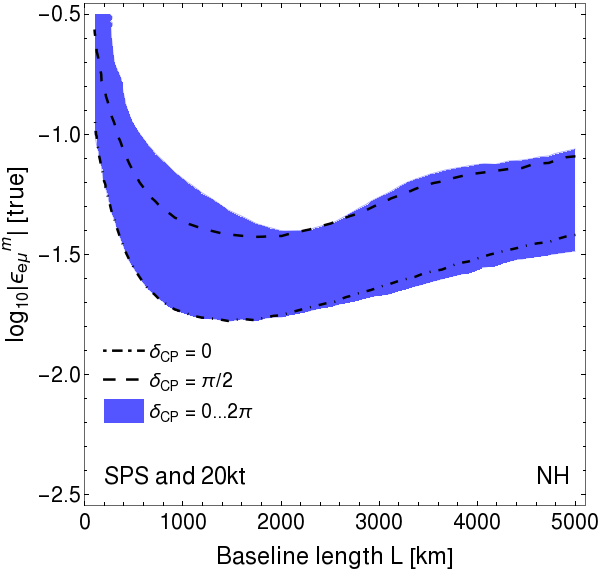}
\caption{Case $\varepsilon^m_{e\mu}$} \label{fig:b}
\end{subfigure}

\medskip
\begin{subfigure}{0.41\textwidth}
\includegraphics[width=\linewidth]{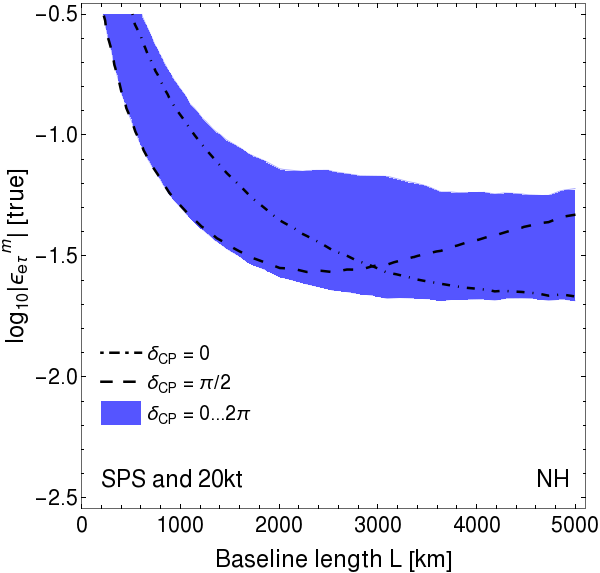}
\caption{Case $\varepsilon^m_{e\tau}$} \label{fig:c}
\end{subfigure}\hspace*{\fill}
\begin{subfigure}{0.41\textwidth}
\includegraphics[width=\linewidth]{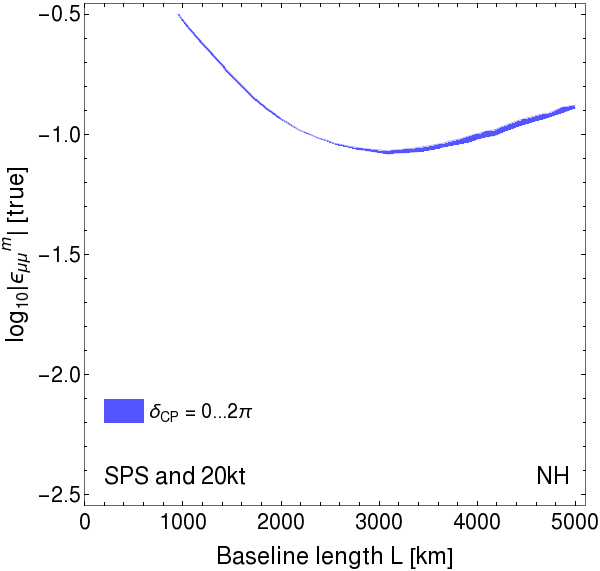}
\caption{Case $\varepsilon^m_{\mu\mu}$} \label{fig:d}
\end{subfigure}

\medskip
\begin{subfigure}{0.41\textwidth}
\includegraphics[width=\linewidth]{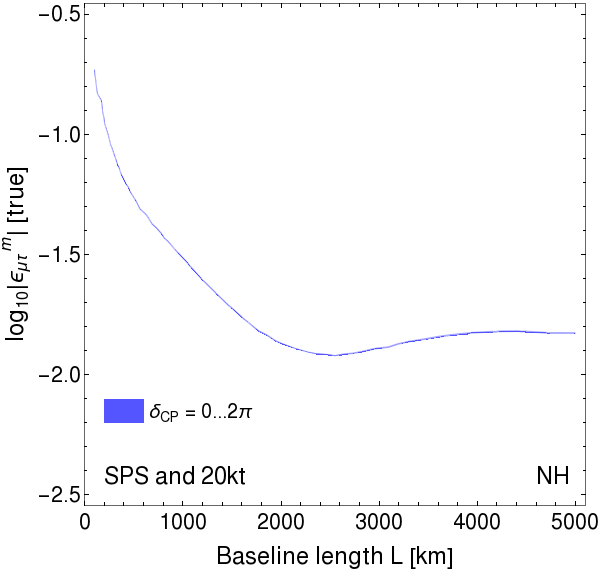}
\caption{Case $\varepsilon^m_{\mu\tau}$} \label{fig:e}
\end{subfigure}\hspace*{\fill}
\begin{subfigure}{0.41\textwidth}
\includegraphics[width=\linewidth]{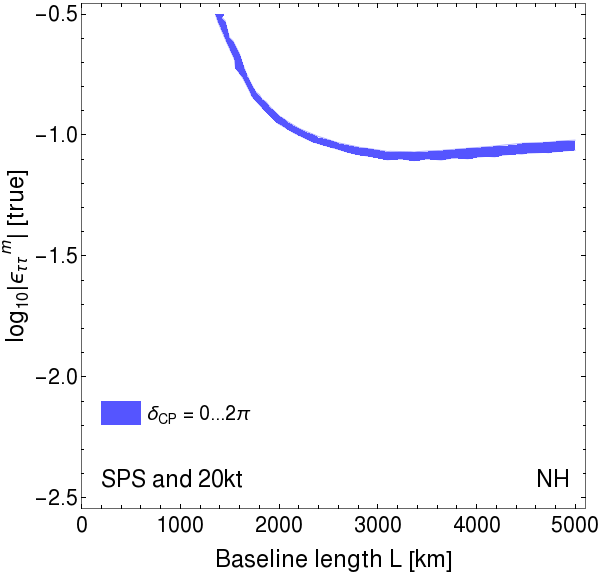}
\caption{Case $\varepsilon^m_{\tau\tau}$} \label{fig:f}
\end{subfigure}
\caption{90 $\%$ confidence limit discovery reach of NSI parameters as a function of baseline length. Band thickness corresponds to the strength of correlation between $\delta_{CP}$ and $\varepsilon^m_{\alpha\beta}$.\label{fig:1}}
\end{figure}

Let us consider the $\nu_{e}\to\nu_{\mu}$ transition. For the amplitude of this process we obtain the expression

\begin{align}
\frac{2E_{\nu}}{L}A_{e\mu} &= Ac_{12}c_{13}(-s_{12}c_{23}-s_{23}s_{13}e^{-i\delta})+ s_{12}c_{13}\Delta m_{21}^2(c_{12}c_{23}-s_{12}s_{13}s_{23}e^{-i\delta}) \notag\\
&+ s_{13}e^{-i\delta}(\Delta m_{31}^2s_{23}c_{13})\notag \\
&+ Ac_{12}c_{13}\varepsilon^m_{e\mu}(c_{12}c_{23}-s_{12}s_{23}s_{13}e^{-i\delta})+As_{12}c_{13}\varepsilon^m_{e\mu}(-s_{12}c_{23}-s_{23}s_{13}e^{-i\delta})+O(s_{13}^2)\notag \\
&\equiv N_{e\mu}+K_{e\mu}e^{-i\delta}+O(s_{13}^2),
\end{align}

where  we have denoted  $s_{ij}\equiv \sin \theta_{ij}$ and $c_{ij} \equiv \cos \theta_{ij}$ ($i,j=1,2,3$). We find the transition probability to have the structure
\begin{equation}
P_{e\mu} \propto N_{e\mu}^2+K_{e\mu}^2+2N_{e\mu}K_{e\mu}\cos \delta.
\end{equation}
The effect of the CP violating angle $\delta$ becomes evident when we evaluate the values of the functions $N_{e\mu}$ and $K_{e\mu}$ for some value of the NSI parameter $\varepsilon^m_{e\mu}$. {\it E.g.} for  $\varepsilon^m_{e\mu} = 10^{-3}$, we get
\begin{equation}
 P_{e\mu} \propto (1.56-1.22\cos\delta)\cdot 10^{-7},
 \end{equation}
It is seen that varying $\delta$ causes large variation in the transition probability $P_{e\mu} $. Thereby it strongly interferes with the determination of the NSI parameter $\varepsilon^m_{e\mu}$. The wide band in the numerical plot of Fig.~\ref{fig:b} is explained, since the plot is obtained by allowing $\delta$ to vary in the range 0 to 2$\pi$. Similarly we find in the case of $P_{e\tau}$ that $P_{e\tau} \propto (1.32 + 1.08\cos \delta)\cdot 10^{-7}$, and again a wide band is caused by variation of $\delta$.

Let us next consider the probability of the transition  $\nu_{\tau}\to\nu_{\tau}$  for comparison. One can show that the dependence on the CP angle is in this case of the form
\begin{equation}
P_{\tau\tau} \propto  N_{\tau\tau}^2+K_{\tau\tau}^2\cos^2\delta-2N_{\tau\tau}K_{\tau\tau}\cos\delta .
\end{equation}
Taking here $\varepsilon^m_{\tau\tau} = 10^{-3}$, we find 
\begin{equation}
P_{\tau\tau} \propto  (3.200+0.120\cos\delta+0.001\cos^2\delta)\cdot 10^{-6}.
\end{equation}
Hence, the transition probability $P_{\tau\tau}$ has significantly smaller dependence on the CP angle than the transition probability  $P_{e\mu}$. This is seen in the discovery reach plot of the NSI parameter $\varepsilon^m_{\tau\tau}$ in Fig.~\ref{fig:f}, where  the CP variation band is much  narrower than the corresponding band in the case of the  $\nu_{e}\to\nu_{\mu}$ transition in Fig.~\ref{fig:b}.

Similarly as in the cases considered above, one can qualitatively understand the CP angle dependence of the discovery reach plots of all the other NSI parameters $\varepsilon^m_{\ell\ell'}$ shown in Fig. 1. As seen in the plots, there are two transitions, where the effect of the CP angle $\delta$ on the determination of the NSI parameters is much larger than in the rest of the transitions, namely the $\nu_{e}\to\nu_{\mu}$ and $\nu_{e}\to\nu_{\tau}$. Note in particular that $K_{ee}(\varepsilon) = 0$, meaning that survival probability $P_{ee}$ is independent of $\delta_{CP}$. 

\begin{figure}[H]
\begin{center}
\includegraphics[scale=0.55]{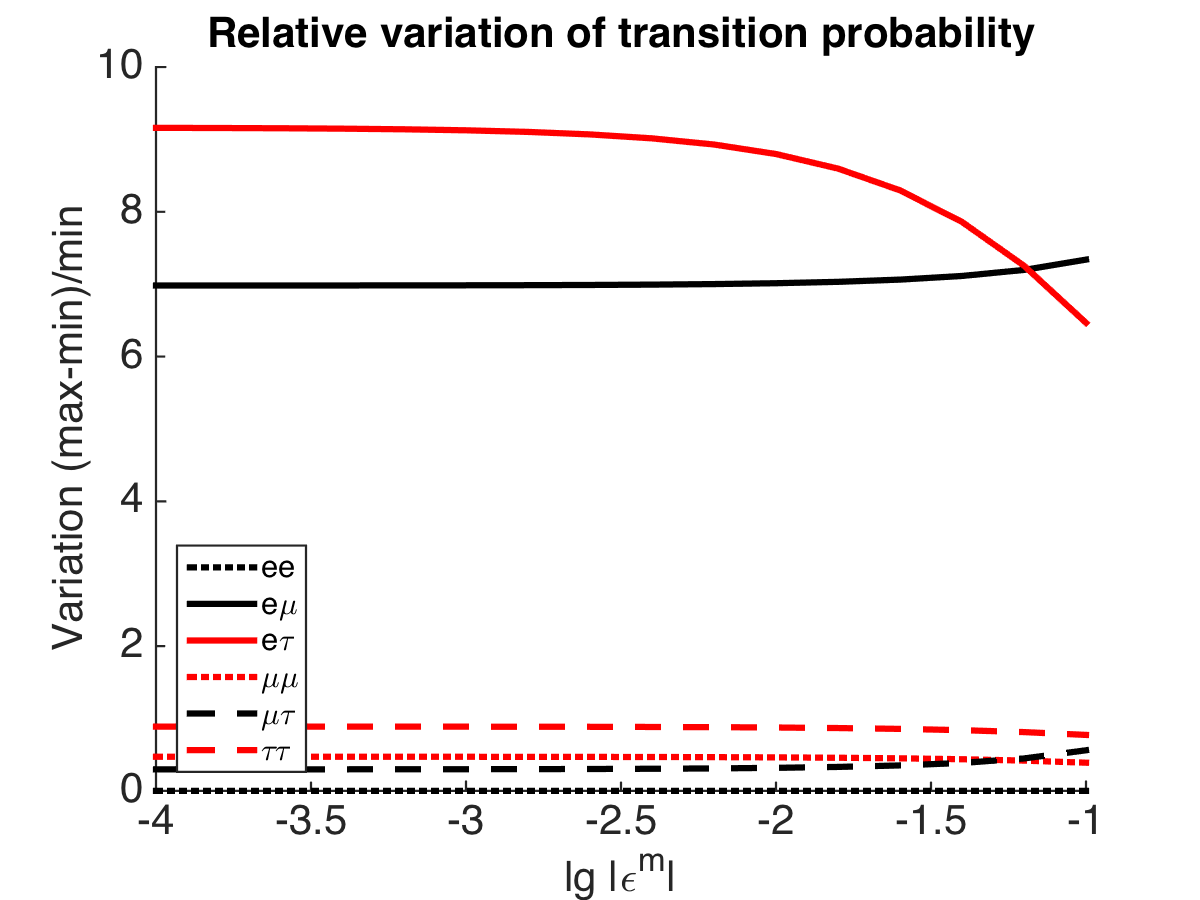}
\end{center}
\caption{\label{fig:2}Relative variation $R$ of six different matter NSI parameter as a function of $\log_{10}|\varepsilon^m_{\ell\ell'}|$. The CP  angle $\delta$ is varied in the range (0,$2\pi$).}
\end{figure}

We have elaborated the effect of the CP angle on the experimental reach of the NSI parameters further in Figure~\ref{fig:2}. There we have plotted the relative variation of the discovery reach as a function of the NSI parameter for different transitions for the CP angle $\delta$ varying in the range 0 to $2\pi$. More precisely, the curves plotted present, as a function of the relevant NSI parameter $\varepsilon_{\ell\ell'}^m$,  the quantities
\begin{equation}
R \equiv  \frac{P_{\ell\ell'}^{\rm max}-P_{\ell\ell'}^{\rm min}}{P_{\ell\ell'}^{\rm min}},
\end{equation}
where $P_{\ell\ell'}^{\rm max}$ and $P_{\ell\ell'}^{\rm min}$ are, respectively, the largest and the smallest value the transition probability achieves when  $\delta_{CP}$ varies in the range 0 to $2\pi$ for a given value of
 $\varepsilon_{\ell\ell'}^m$. The magnitude of the variation is approximately constant when $\log_{10}|\varepsilon^m| < -2.5$ for all parameters. For an example value of $\varepsilon^m_{\ell\ell'}$, the variation with respect to CP angle as a function of energy was considered in \cite{Shoemaker:2012,Mehta:2015}.

 \pagebreak
 \thispagestyle{empty}

\pagebreak

\begin{table}
\begin{center}
\begin{tabular}{|c|l|}\hline
\rule{0pt}{3ex}\textbf{Function} & \textbf{Expression}\\ \hline
\rule{0pt}{3ex}$K_{e\mu}$ & $(-Ac_{12}c_{13}s_{23}-s_{12}^2c_{13}\Delta m_{21}^2s_{23}+\Delta m_{31}^2s_{23}c_{13}-Ac_{13}\varepsilon^m_{e\mu}(c_{12}+1)s_{12}s_{23})s_{13}$ \\ \hline
\rule{0pt}{3ex}$N_{e\mu}$ & $-Ac_{12}c_{13}s_{12}c_{23}+s_{12}c_{13}\Delta m_{21}^2c_{12}c_{23}+Ac_{13}c_{23}\varepsilon^m_{e\mu}(c_{12}^2-s_{12}^2)$\\ \hline 
\rule{0pt}{3ex}$K_{ee}$ & 0\\ \hline
\rule{0pt}{3ex}$N_{ee}$ &$A(1+\varepsilon^m_{ee})c_{12}^2c_{13}^2+\Delta m_{21}^2s_{12}^2c_{13}^2+\Delta m_{31}^2 s_{13}^2$\\ \hline
\rule{0pt}{3ex}$K_{e\tau}$ &$(-Ac_{12}^2c_{13}c_{23}-\Delta m_{21}^2s_{12}^2c_{13}s_{23}+A\varepsilon^m_{e\tau}s_{13}s_{23}+c_{23}c_{13}\Delta m_{31}^2)s_{13}$\\ \hline
\rule{0pt}{3ex}$N_{e\tau}$ & $Ac_{12}c_{13}(s_{12}s_{23}+\varepsilon^m_{e\tau}c_{23}c_{13})-\Delta m_{21} ^2s_{12}c_{13}c_{12}c_{23}$\\ \hline
\rule{0pt}{3ex}$K_{\mu\mu}$ & $2s_{12}s_{13}s_{23}c_{12}c_{23}((1-\varepsilon^m_{\mu\mu})A-\Delta m_{21}^2)$\\ \hline
\rule{0pt}{3ex}$N_{\mu\mu}$ & $A(s_{12}^2c_{23}^2+c_{12}^2s_{23}^2s_{13}^2)+(\Delta m_{21}^2+A\varepsilon^m_{\mu\mu})(c_{12}^2c_{23}^2+s_{12}^2s_{23}^2s_{13}^2)+\Delta m_{31}^2s_{23}^2c_{13}^2$ \\ \hline
\rule{0pt}{3ex}$K_{\tau\tau}$ & $\Delta m_{21}^2(c_{12}^2s_{23}^2+s_{12}^2c_{23}^2s_{13}^2+2c_{12}s_{23}s_{12}c_{23}s_{13})-2As_{12}s_{23}c_{12}c_{23}s_{13}$\\ \hline
\rule{0pt}{3ex}$N_{\tau\tau}$ & $A(s_{12}^2s_{23}^2+c_{12}^2c_{23}^2s_{13}^2)+(\Delta m_{31}^2+A\varepsilon^m_{\tau\tau})c_{23}c_{13}$\\ \hline
\rule{0pt}{3ex}$N_{\mu\tau}$ & $-s_{12}^2s_{23}c_{23}A+\Delta m_{21}^2s_{23}c_{23}(-c_{12}^2+s_{12}^2s_{13}^2)+s_{23}c_{13}^2\Delta m_{31}^2c_{23}+A\varepsilon^m_{\mu\tau} c_{12}c_{13}(c_{23}^2-s_{23}^2)$\\ \hline
\rule{0pt}{3ex}$K_{\mu\tau}^{(c)}$ & $-4A\varepsilon^m_{\mu\tau} s_{12}s_{13}s_{23}c_{13}c_{23}+2s_{23}^2\Delta m_{21}^2s_{12}s_{13}c_{12}-2s_{23}^2As_{12}s_{13}c_{12}$\\ \hline
\rule{0pt}{3ex}$K_{\mu\tau}^{(-)}$ & $-\Delta m_{21}^2s_{12}s_{13}c_{12}+As_{12}s_{13}c_{12}$\\ \hline
\end{tabular}
\caption[]{\label{NK}
The expressions of the functions $K_{\ell\ell'}$ and $N_{\ell\ell'}$, where $\varepsilon^m_{\ell\ell'}\ll 1$ is assumed.}
\end{center}
\end{table} 
 
\section{Summary}

We have investigated the prospects of long baseline neutrino experiments to probe the possible existence of the non-standard neutrino interactions, or NSI, using the LBNO setup as our benchmark. We have neglected the effects of NSI at the source or detector and taken into account only the effects during neutrinos traveling in the Earth's crust. This is justified, because the bounds on the source and detector NSI are about an order of magnitude stricter than on matter NSI \cite{Biggio:2009nt,Liao:2016hsa}. It is found that the discovery sensitivity depends on the baseline length, the best sensitivity for this experimental setup being achievable at about 2000 km. We also found that the CP angle $\delta$ quite severely interferes with the NSI search in the $\nu_{\mu}\leftrightarrow\nu_{e}$ and $\nu_{\tau}\leftrightarrow\nu_{e}$ transitions, while in all the other transitions the effect is negligible. The most sensitive probe would be the $\nu_{\mu}\leftrightarrow\nu_{e}$ and  $\nu_{\mu}\leftrightarrow\nu_{\tau}$ transitions. In the former channel one would be able to limit the value of the NSI parameter $\varepsilon^m$ below 0.02, if the CP angle is close to $\delta_{CP} =\pi/2$, for other values of $\delta$ the achievable bound is less stringent. In the case of  the $\nu_{\mu}\leftrightarrow\nu_{\tau}$  channel the sensitivity is 0.01 independently of the value of $\delta$. In both cases the most stringent limit requires the baseline to exceed 2000 km. 

The role of baseline length in studying matter NSI effects can be demonstrated by comparing the potential of T2HK \cite{Abe:2015zbg} and DUNE \cite{Acciarri:2015uup} whose abilities to constrain $\varepsilon^m$ have been studied in \cite{Liao:2016hsa,Bass:2013vcg,Coloma:2015kiu,deGouvea:2015ndi}. It is shown in these studies that the 295-km-long T2HK experiment provides significantly lower sensitivity to constrain $\varepsilon_{\alpha\beta}^m$ than the 1300-km-long DUNE project. 
Similarly, the benchmark setup used in this work with 2300-km-long baseline improves the sensitivity compared to DUNE even with the 20 kt detector mass.
Comparing the baseline length using otherwise our benchmark setup shows that for each NSI parameter, there is an optimal baseline length which also depends on the CP violating phase.

\section{Acknowledgments}
\begin{acknowledgments}
KH acknowledges the H2020-MSCA-RICE-2014 grant no. 645722 (NonMinimalHiggs). TK expresses his gratitude to the Magnus Ehrnrooth foundation for financial support.
\end{acknowledgments}
\pagebreak

\bibliography{bibliography}

\end{document}